\documentclass[journal,transmag]{IEEEtran}
\ifCLASSINFOpdf
\else
\fi

\usepackage{booktabs} 
\usepackage{amsmath} 
\usepackage{graphicx} 
\usepackage{algorithm} 
\usepackage{algpseudocode}
\usepackage{algorithmicx}
\usepackage{subfigure} 
\usepackage{calc}
\usepackage{float}
\usepackage{lipsum}

\usepackage[ 
n,
advantage,
operators,
sets,
adversary,
landau,
probability,
notions,
logic,
ff,
mm,
primitives,
events,
complexity,
asymptotics,
keys
] {cryptocode}

\makeatletter
\newenvironment{breakablealgorithm}
{
	\begin{center}
		\refstepcounter{algorithm}
		\hrule height.8pt depth0pt \kern2pt
		\renewcommand{\caption}[2][\relax]{
			{\raggedright\textbf{\ALG@name~\thealgorithm} ##2\par}%
			\ifx\relax##1\relax 
			\addcontentsline{loa}{algorithm}{\protect\numberline{\thealgorithm}##2}%
			\else 
			\addcontentsline{loa}{algorithm}{\protect\numberline{\thealgorithm}##1}%
			\fi
			\kern2pt\hrule\kern2pt
		}
	}{
		\kern2pt\hrule\relax
	\end{center}
}
\makeatother

\begin{document}
%
\title{NIMSA: Non-Interactive Multihoming Security Authentication Scheme for vehicular communications in Mobile Heterogeneous Networks\\


\thanks{Ping Dong\IEEEauthorrefmark{*} is the corresponding author.}

}



\author{\IEEEauthorblockN{Zongzheng Wang${^1}$, Ping Dong${^1}$\IEEEauthorrefmark{*}}
	\IEEEauthorblockA{${^1}$School of Electronic and Information Engineering, Beijing Jiaotong University, P. R. China\\
		Email: \{18111035, pdong\} @bjtu.edu.cn
	}
	
}

%



\IEEEtitleabstractindextext{%
\begin{abstract}


In vehicular communications, in-vehicle devices' mobile and multihoming characteristics bring new requirements for device security authentication. On the one hand, the existing network layer authentication methods rely on the PKI system; on the other hand, key negotiation needs interaction. These two points determine that the traditional security authentication method requires bandwidth consumption and additional delay. It is unsuitable for heterogeneous wireless scenarios with a high packet loss rate and limited bandwidth resources. In addition, the establishment of a security association state is contrary to the original design that the network layer only provides a forwarding function. We proposed a non-interactive multihoming security authentication (NIMSA) scheme, a stateless network layer security authentication scheme triggered by data forwarding. Our scheme adopts an identity-based non-interactive key agreement strategy to avoid the interaction of signaling information, which is lightweight and has good support for mobile and multipath parallel transmission scenarios. The comparison with IKEv2 and its mobility and multihoming extension scheme (MOBIKE) shows that the proposed scheme has shorter authentication and handover delay and data transmission delay and can bring better bandwidth aggregation effect in the scenario of multipath parallel transmission.
\end{abstract}

}

\maketitle

\IEEEdisplaynontitleabstractindextext

%
\IEEEpeerreviewmaketitle

\section{Introduction}
%
%
%
%


\IEEEPARstart{A}{s} the development trend of future transportation, intelligent transportation system (ITS) is a hot topic \cite{sumalee2018smarter}. With the development of autonomous driving in ITS, In-vehicle devices will generate a large amount of privacy-related user data, such as location, speed, and user application data, which need to be safely transmitted through vehicles-to-infrastructure (V2I) \cite{srinivas2020designing}. With the maturity of 5G technology and the development of future 6G technology, wireless communication technology has brought multiple heterogeneous network resources. In-vehicle devices usually have multihoming interfaces for accessing multiple heterogeneous networks \cite{camacho2018emerging}. The multihoming feature can effectively prevent the single point failure of the mobile router and the connection interruption caused by mobile handoff. In addition, the multihoming feature brings the opportunity to make full use of heterogeneous network resources in the mobile process. The characteristics of multihoming bring many advantages but also many security challenges. In the wireless heterogeneous network scenario, the channel is vulnerable to passive eavesdropping, signal active interference, and interference similar to DDoS attacks. Secondly, continuous and unpredictable mobility (multihoming can be seen as another manifestation of mobility) blurs the difference between normal and abnormal, which is easy to cause source IP address spoofing. In addition, the link state of heterogeneous wireless networks fluctuates heterogeneously. The horizontal and vertical switching brought about by the highly dynamic and multihoming characteristics of in-vehicle devices poses challenges for end-to-end data source authentication from in-vehicle devices to infrastructure \cite{2021Secure}.  

This paper focuses on the vehicle to infrastructure safety communications scenario in which the vehicle has multihoming features. We ensure the communication continuity of the proposed communication model by using the multihoming expansion of network mobility basic support (NEMO) \cite{tsirtsis2011flow}. In the V2I communication process, the In-vehicle device (also known as Mobile Router, MR), as the traffic export of the vehicle subnet, integrates a variety of heterogeneous communication technologies, such as 5G, LTE, WiFi, and Satellite-link. MR establishes a connection with the ground infrastructure (also known as HOME Agent, HA) by obtaining multiple care-of addresses. The core of the security mechanism is to achieve authentication and key negotiation for frequently switched multiple care-of addresses. In response to this demand, there are mainly two types of research directions, IPsec/ike, and its mobility and multihoming extension (MOBIKE) \cite{eronen2006ikev2}, and host identity protocol (HIP) \cite{moskowitz2015host} that natively supports mobile multihoming. However, whether it is the MOBIKE extension or the HIP that naturally supports mobility and multihoming, the core of its construction of secure authentication communication is still the establishment of the Security Associations (SA); that is, the authentication of device identity is realized through the traditional PKI certificate mode between end-to-end, and the exchange of critical materials is realized through the interactive key exchange protocol based on Diffie-Hellman algorithm, At the same time, for IP handoff caused by mobility as well as the support of multihoming is realized by way of control signaling.

Moreover, the existing security strategies lack research on support for concurrent multipath transmission. Although there are related extensions to multipath, it isn't easy to make full use of complex and heterogeneous network resources because the original design did not consider multipath concurrency. Especially in the multihomed NEMO scenario, the role of MR is to provide the aggregation and forwarding function of data packets in the network layer. However, the establishment of SA makes the stateless network forwarding establish relevant states, which violates the functional characteristics of the network layer. Therefore, in the high-speed mobile heterogeneous network scenario, this security authentication measure based on Security Associations shows low efficiency, high cost, and inflexibility. These drawbacks are precisely intolerable for latency-sensitive vehicle applications in ITS scenarios.

We summarize the problems of existing authentication schemes as follows:

\begin{itemize}
\item Device authentication relies on the complex PKI system, and the certificate brings the overhead problem. Its online certificate verification or CRL (Certificate Revocation List) introduces latency and consumes bandwidth.
\item The establishment of SA requires the interaction of control signaling, which brings authentication delay and breaks the stateless forwarding characteristic of the IP layer. Once SA changes state (such as the handover caused by IP address movement), it needs the interaction of control signaling to realize the change or even re-establishment of security alliance state, and its signaling interaction will inevitably cause communication delay, In the wireless heterogeneous network scenario, link delay and retransmission of establishment information caused by high packet loss rate will bring significant authentication delay.
\item No matter HIP or IKEv2, the support for multihoming only stays in the form of providing a spare address list to use multihoming interface as primary and standby communication, and the support for parallel transmission is insufficient.
\end{itemize}

Given the above problems, we propose NIMSA, an identity-based non-interactive multihoming security authentication scheme for mobile multihoming devices which works in high-speed mobile heterogeneous network scenarios and has adapter interfaces for accessing multiple network resources. NIMSA scheme is based on the bilinear pairing system; The device generates the data authentication key through the pre-allocated private key, device ID, network adapter serial number, and IP address information. The identity-based non-interactive key negotiation makes the interaction of authentication and key interaction information only included in the data transmission header without establishing the security state in the mobile multihoming scenario. compared with the traditional strategy; it has the characteristics of lightweight and high efficiency.

The contributions of this article are as follows.
\begin{itemize}
\item We propose an identity-based non-interactive multihoming security authentication scheme, which avoids the overhead in traditional certificate authentication and the communication delay with certificate authority (CA). At the same time, the non-interactive key negotiation avoids the transmission delay caused by key information interaction.
\item We designed the device security authentication and authentication switching process in the mobile heterogeneous network and realized the multi-homed support for the device through the header design. Due to the data-driven authentication establishment and mobile switching mechanism, we adhered to the stateless forwarding principle of the IP layer and improved the system's flexibility.
\item Through simulation comparison, we prove the advantages of the proposed NIMSA scheme compared with the existing strategies in authentication establishment and handover delay, data transmission delay, and multipath aggregation transmission support.
\end{itemize}

The rest of this paper is organized as follows. We introduce the related research in Section II. Section III analyzes the security authentication requirements in the mobile multihoming scenario combined with the existing research strategies. In Section IV, we provide the detailed design of the NIMSA scheme. We design experiments to analyze the performance of the proposed scheme in Section V. Finally, we conclude this article in Section VI.


\section{Related Work}


In this section, we introduce the related work. Our V2I security model is based on NEMO to provide support. We first introduced related research on NEMO and its multihoming expansion in the vehicle network. We reviewed its mobile multihoming security research. Then we present the relevant research on the identity-based non-interactive negotiation technology used in this paper.

\subsection{Multihoming extension of NEMO}

With the growth of network access demand of mass devices in mobile scenarios, the IETF network mobility Working Group proposed NEMO Basic Support (NEMO-BSP), NEMO-BSP ensures session continuity for all the nodes in the Mobile Network, even as the Mobile Router changes its point of attachment to the Internet \cite{devarapalli2005network}. Many studies have optimized NEMO \cite{islam2020design}. To make full use of heterogeneous network resources and reduce the switching overhead, IETF proposed the multihoming extension of NEMO-BSP, \cite{tsirtsis2011flow}\cite{2015Flow} introduces multihoming using mobile IP and schemes with IP Flow Bindings. There have been related researches on the multipath transmission of network-layer devices. NEMO supporting multihoming can be regarded as an intermediate network layer device that uses the proxy to realize multipath transmission \cite{2004MAR} \cite{li2016multipath}. In this paper, we adopt the multipath communication architecture proposed in the related work\cite{2016Improving} to realize the multihoming communication scenario of related NEMO. Given the security authentication problem between MR and HA in NEMO, there are two main research ideas, using IKEV2 and HIP protocol to realize the security deployment of mobile multihoming. Next, we will introduce them respectively.


\subsubsection{IKEv2 and MOBIKE extension for Multihoming NEMO}

As mentioned above, the standard PMIPv6 protocol only supports network-based mobility for single nodes. Therefore, to extend the support of PMIPv6 to mobile networks, Lee et al. introduced a P-NEMO scheme \cite{islam2020design} based on PMIPv6. In P-NEMO, an onboard router, known as an MR, receives a mobile network prefix (MNP) and a home network prefix (HNP), which can be attributed to an extension of the binding update lists located at the infrastructure entities, namely, the MAG and the LMA. With the MNP, the local moving network served by the MR is enabled with IP mobility support. The P-NEMO scheme aims to reduce the signaling load while maintaining Internet connectivity for the moving networks.

In NEMO-BPS, IPSec is used by default to protect the secure communication from MR to HA. The security authentication and SA establishment depend on the IKE protocol. IKEv2 is the latest version of the Internet Key Exchange Protocol, which performs mutual authentication between two parties and establishes an IKE Security Association (SA) that includes shared secret information that can be used to establish IPsec SAs \cite{2010Internet} efficiently. To expand the support for mobile multihoming devices, IETF IKEv2 Mobility and multihoming (MOBIKE) working group developed IKEv2. This extension allows clients who have established IPSec SA to realize address translation without breaking the connection and reestablishing IPSec SA. \cite{2011End} studied the nested security between MN and CN in NEMO based on IPSec, but the multihoming scenario is not considered. Fernandez et al. studied the mobility, and seamless inter-technology handovers of IPv6 VANET in heterogeneous networks \cite{2016Mobility} \cite{Fernandez2017Towards}, and proposed a security approach using IKEv2 and IPsec. Their approach focuses on using MR with multiple wireless interfaces to secure IPv6 NEMO for internal vehicle devices, and the wireless interfaces consist of multiple heterogeneous links\cite{Pedro2016Securing}. However, in their research, multipath is only used to transmit different streams and does not support parallel transmission. In \cite{2017Achieving} the seamless communication in software-defined vehicular networks is studied, and its security mechanism is still a variant of IKEv2 and IPSec. For resource-constrained scenarios, \cite{2020TAKE} proposed a lightweight authenticated key agreement protocol based on IPSec. \cite{2021Secure} summarizes the IP-based security mechanism in vehicular networks, a typical scenario in NEMO.


\subsubsection{Host Identity Protocol for Multihoming NEMO}

Host identity protocol \cite{moskowitz2015host} is an internetworking architecture that adds a namespace between the IP layer and the transport protocols. In this namespace, hosts are identified by Host Identifiers (HI), to which application-layer protocols are bound instead of IP addresses. In this way, the problem of semantic overload of IP addresses is avoided. In other words, it realizes the separation of identity and location in the general sense. Thereby enabling continuity of communications across IP address changes \cite{rfc8046}. At the same time, the same HI can be mapped to multiple IP addresses to realize multi-host communication\cite{rfc8047}. The nature of HIP is very suitable for the multihoming NEMO scene. \cite{novaczki2008design} proposed HIP-NEMO and carried out a series of performances. \cite{toledo2011analytical} evaluated The performance improvement of hip registration for Nemo. \cite{toledo2013design} improve the security of hip-NEMO and ensure the security through additional authentication DH exchange. C-HIP is proposed to study key agreement in resource-constrained scenarios in \cite{2017CHIP}. An authentication method based on the implicit certificate is designed for lightweight requirements in \cite{hossain2020p}. Unfortunately, HIP based security scheme is similar to IKE, which relies on the certificate system, and the secure communication depends upon the establishment of SA. In addition, hip is a layer 3.5 protocol based on IP. It is an end-to-end protocol and needs to change the protocol stack, so it isn't easy to deploy.


\subsection{Identity-based non-interactive key exchange}

The identity-based cryptographic system was introduced by Shamir in 1984, in which the user's public key is extracted by identity to avoid the cost of certificate authentication brought by traditional PKI system \cite{shamir1984identity}. Based on an identity-based cryptosystem, Identity-based non-interactive key distribution (ID-NIKD) is proposed, a cryptographic primitive that enables two parties to negotiate a shared secret key without exchanging key materials. The first two-party ID-NIKD scheme is proposed by Sakai et al. \cite{sakai2000cryptosystems}, namely the SOK scheme. The following related research is based on the SOK scheme, and the security extension and related proof are carried out \cite{dupont2006provably}\cite{paterson2009relations}\cite{chen2016sakai}. The security of the algorithm is based on the bilinear pairing problem on the elliptic curve. There are many researches on the application of SOK algorithm. \cite{schridde2009trueip} studied the IP address signature authentication protocol used to prevent IP spoofing attacks. in \cite{wu2013non}, Non-interactive key agreement is applied to devise authentication. \cite{wang2018t} developed a network layer security protocol based on IP address self-authentication, which is suitable for secure authentication communication in the local area network. \cite{dharminder2021edge} takes the same idea as \cite{wang2018t}, the authors apply the non-interactive key agreement to vehicular communications and build a communication system that does not require a trusted party.


Most of the application scenarios of these studies are small-scale networks, and the communication process can be abstracted into the following steps:

\begin{itemize}
	\item The private key generator (PKG) establishes the system and generates system parameters and master keys.
	\item The communication participant registers and obtains its private communication key from PKG.
	\item The communication entity forms a shared key to communicate through the peer ID and its private key.
\end{itemize}

In this paper, to realize the secure communication of mobile multi-homing, combined with the classic SOK algorithm, We transform the traditional security mechanism to adapt to the mobile multi-homed environment.



\section{Authentication requirement analysis in mobile multihoming scenario}

\begin{figure}[!t]
	\centering
	\includegraphics[width=3.5in]{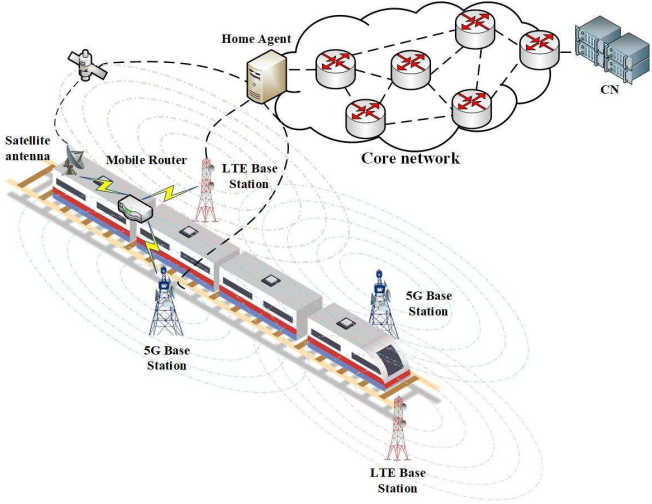}
	\caption{High speed mobile multihoming scene}
	\label{fig3.1}
\end{figure}

This paper aims at the authentication of multihoming NEMO in high-speed mobile scene, as shown in the Fig.\ref{fig3.1}. In this scenario, the wireless link resource status is unstable, with time-varying characteristics, high packet loss rate, and limited network bandwidth resources. The main secure authentication transmission methods in the existing research are IKE-based IPSec and HIP. Next, we will summarize the two ways in the process of authentication and switching.

\subsection{Authentication flow}

\begin{figure}[!t]
	\centering
	\includegraphics[width=2.8in]{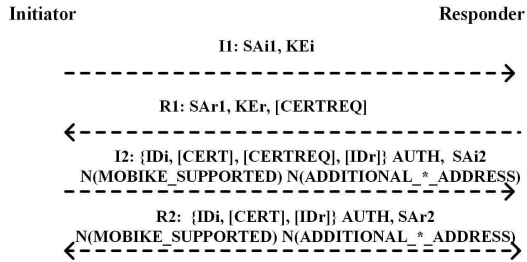}
	\caption{the IKEv2 Initial Exchanges}
	\label{IKE-AUTH}
\end{figure}

\begin{figure}[!t]
	\centering
	\includegraphics[width=2.8in]{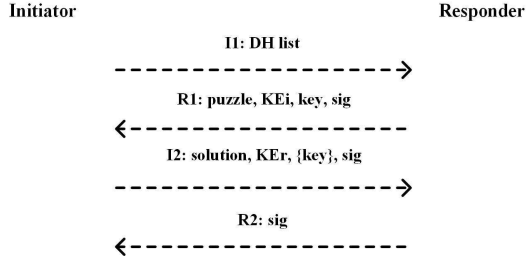}
	\caption{the HIP Base Exchange}
	\label{HIP-AUTH}
\end{figure}

Fig \ref{IKE-AUTH} shows that the authentication establishment process in IKEv2 needs to go through the DH algorithm of exchanging key materials and certificate exchange for authentication. It is worth mentioning that mobike extension needs to support information interaction and provide an address list for multihoming support.

As can be seen from Fig \ref{HIP-AUTH}, DH exchange with the puzzle is required in the HIP to realize the establishment of the shared key. The key and signature represented by HI are used for authentication in the exchange process. Since I1 and R2 packets carry relevant information, two exchange rounds are also required.

\subsection{switching flow}

\begin{figure}[H]
	\centering
	\includegraphics[width=2.8in]{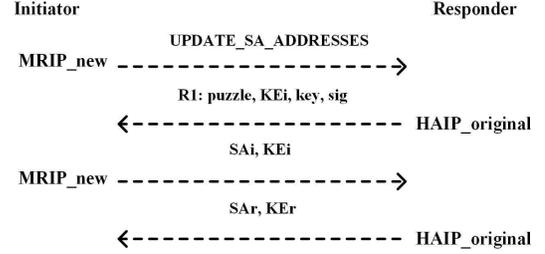}
	\caption{IKEv2 MOBIKE extension}
	\label{IKE-HANDOFF}
\end{figure}
\begin{figure}[H]
	\centering
	\includegraphics[width=2.8in]{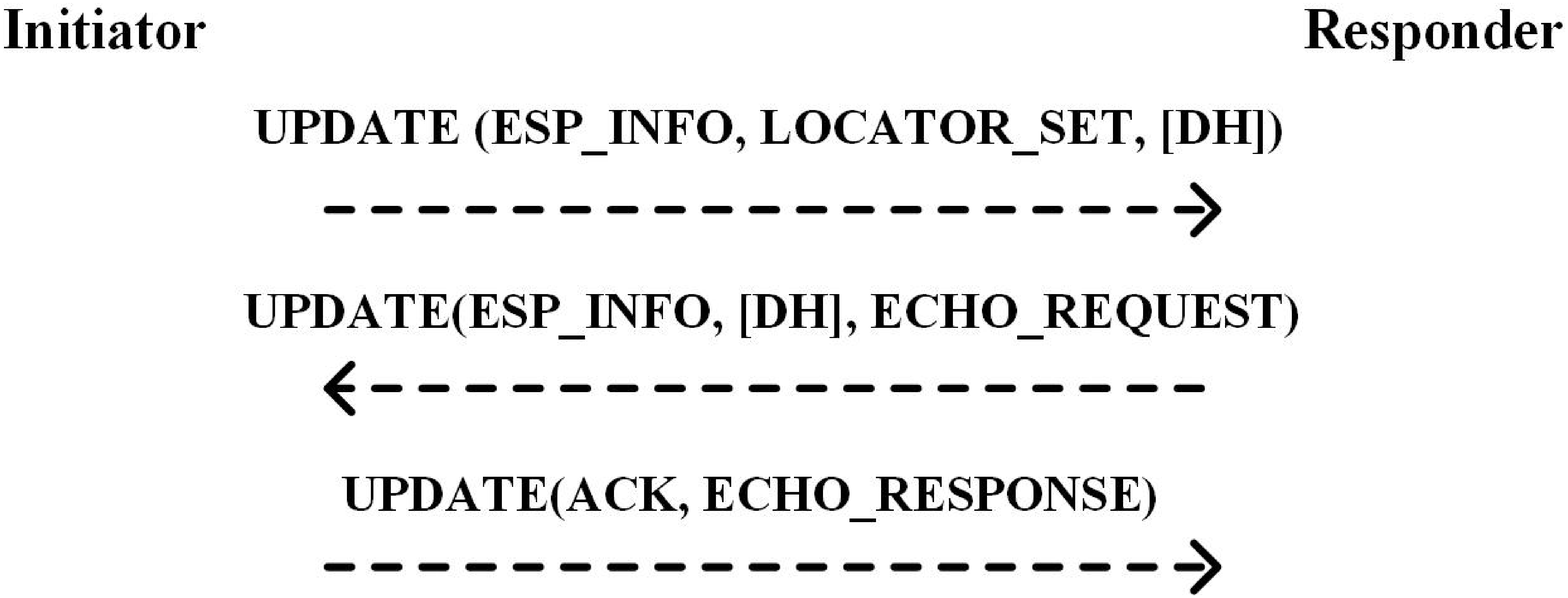}
	\caption{HIP Mobility and Multihoming}
	\label{HIP-HANDOFF}
\end{figure}

In the handover process,  Fig \ref{IKE-HANDOFF} and \ref{HIP-HANDOFF} show that the handover initiator needs to send an address update notification. IKE needs to send a UPDATA\_SA\_ADDRESSES message, and HIP sends updated LOCATOR\_SET information. In contrast, if IKE needs another round of interaction to renegotiate the key for the new address, HIP can send the DH interaction together with the address update notification.

%
%


Through the above analysis, we can see that in the authentication process, IKE and HIP protocol need two rounds of exchange to establish the associated authentication state. The primary purpose is to realize identity authentication and shared key negotiation. In the handover process, the exchange, including notification information and key renegotiation, needs to be sent. However, in the scenario of limited network resources and a high packet loss rate, this information exchange will consume bandwidth resources and bring a long retransmission waiting delay caused by packet loss. Since MR is equivalent to the functional requirements of three-layer forwarding equipment, establishing a state will bring additional performance overhead, so it is necessary to design a lightweight security authentication strategy suitable for the network layer. For the convenience of description, this paper only describes the scenario where one HA corresponds to multiple MR. the expansion principle of the system is the same as that of a single HA scenario. Next, we will introduce the security assumptions.

\begin{figure*}[!t]
	\begin{center} 
		\includegraphics[width=\textwidth]{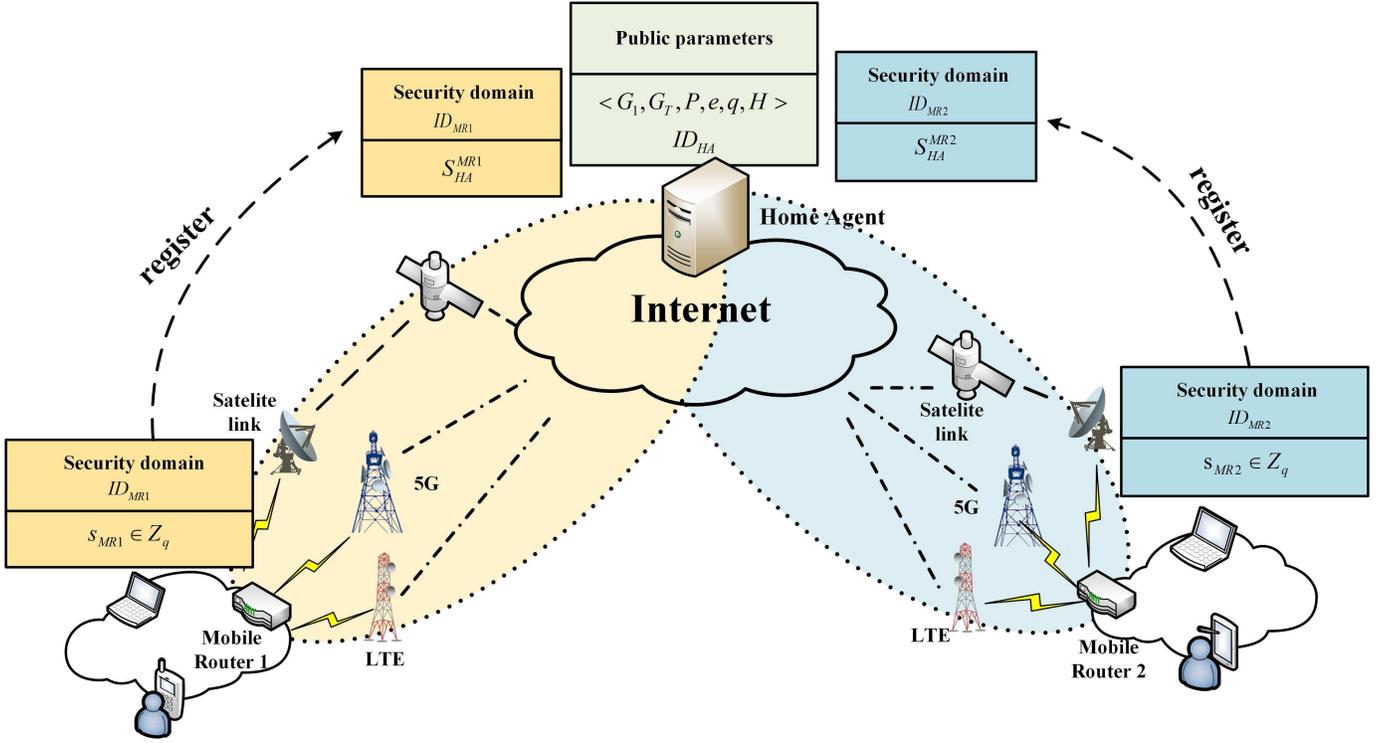} 
		\caption{System establishment and registration} 
		\label{fig3.2}
	\end{center}   
	
\end{figure*}

\subsection{Security assumptions}

Our security system is based on the following security assumptions. We assume that the communication process's mobile router and HA equipment are secure. The mobile router will not communicate as an attacker endangering the system security. When it is known through external factors that the mobile router is no longer secure (such as being controlled and stolen by an attacker, etc.), The security parameters of the router are unregistered in the form of security revocation.

\section{The Design of NIMSA scheme}

\subsection{Security principle}
The construction of the security domain is an identity-based non-interactive key agreement. We will introduce its mathematical principle below:

Pairings and the BDH problem

In a bilinear group system, we define $G_{1}$ as an additive group with prime order $q$, and we define $G_{T}$ as a multiplicative group with prime order $q$. we define $e:G_{1}\times G_{1} \rightarrow G_{T}$ as bilinear map. The map satisfies the following properties:

a) Bilinearity: $\forall$$P,Q$$\in$$G_{1}$, $\forall$$a,b$$\in$$Z_{q}$ we have $e(aP,bQ)=e(P,Q)$.

b) Non-degeneracy: if $P$ is a generator of $G_{1}$ then $e(P,P)$ is a generator of $G_{T}$.

c) Computability: given $P$$\in$$G_{1}$ and $Q$$\in$$G_{1}$, $e(P,Q)$ can be efficiently computed.

The security of the key interactive negotiation protocol shown below depends on the bilinear Diffie-Hellman problem:

Given a generator algorithm PairingGen as just described.

for a probabilistic polynomial-time algorithm $\adv$, we define the advantage of $\adv$ in solving the Bilinear Diffe-Hellman (BDH) problem as

\begin{equation} 
\advantage{bdh}{\adv} = Pr(\adv(aP,bP,cP)=e(P,P)^{abc})
\end{equation}
where $a,b,c \leftarrow Z_q^*$ are chosen uniformly at random. if $\advantage{bdh}{\adv}$ is negligible for all probabilistic polynomial time algorithms $\adv$, then the BDH assumption holds for ($G_{1}$,$G_{T},e$).

\subsection{System establishment}

\textbf{System setup}: Home agent use a probabilistic polynomial time algorithm to generate a bilinear group system, which includes two cycle groups $G_{1}$ and $G_{T}$ with prime order $q$, a bilinear map $e:G_{1}\times G_{1} \rightarrow G_{T}$. Meanwhile, we choose a collision resistant hash function $H:{0,1}^*\rightarrow G_{1}$, and public parameters $<p,G_{1},G_{T},e,H>$ are disclosed.The security domain parameters of the devices associated with this HA are built on these public parameters.The equipment in the system shall have unique identification, expressed as $ID_{HA}$, $ID_{MR}$

\subsection{Secure domain key generation and MR registration}

\textbf{Registration phase}: The mobile router selects $s_{MR}\in$$Z_{q}$ as the primary private key, authenticates identity to HA, and submits registration. The IP address of HA is fixed. Therefore, $S_{HA}^{MR}=s_{MR}H(ID_{HA}||IP_{HA})$ is calculated as the corresponding private key of the HA corresponding to the registered MR. HA records the registered $ID_{MR}$ and saves the corresponding private key $S_{HA}^{MR}$ as the registration voucher of the MR. MR calculates the shared key through $s_{MR}$, identity, address, and relevant information of HA, and HA calculates the shared key through relevant information of MR and $S_{HA}^{MR}$, to realize secure communication. We can regard MR and HA as in the same security domain. The registration process is shown in Fig \ref{fig3.2}.

Different from the traditional application scenario of IBE, the mobile router itself acts as part of the PKG function. According to the idea of \cite{balfanz2003secret}, the registered MR can achieve the purpose of identity authentication. Next, we will introduce the communication process in detail.

\subsection{Key negotiation and source address authentication}

As mentioned above, the existing network layer security strategy depends on the establishment of the state. Before data interaction between communication parties, the exchange of state establishment information is required, and a large amount of delay is introduced in heterogeneous wireless scenarios. Therefore, the security strategy we propose is a data-driven stateless protocol, and When both communication parties have data transmission requirements, they do not need to establish a state similar to the security association first. In this paper, the identity-based non-interactive key is used to realize the security authentication of multi-homing mobile devices. In this design, MR and HA together constitute a security domain. The address identification communication between devices is registering the corresponding address private key with the security center. Through this design, the support for an address change is realized. The communication between devices must specify the security domain of both parties through the form name of device ID transmission in the data. The negotiation of shared keys between devices depends on the device address identification. The specific process is as follows:

Different from the traditional application scenario of IBE, the mobile router itself acts as part of the PKG function. According to the idea of \cite{balfanz2003secret}, the registered MR can achieve the purpose of identity authentication. Next, we will introduce the communication process in detail.

\subsubsection{MR send}

When MR obtains the care-of address in the field and attempts to access HA for data interaction, The network adapter dynamically detects its network status, gets the local IP address and network adapter number $IF_{num}$, and then applies for the corresponding network adapter private key from the MR security domain. $S_{MR}^{IF_{num}}=s_{MR}H(ID_{MR}||IP_{MR}^{IF_{num}}||IF_{num})$. After the sending policy scheduling, the sending network card is selected at the sending end, and the shared key material is calculated according to the destination address and HA identification, which can be described as $K^{IF_{num}}_{MR}=e(S_{MR}^{IF_{num}}, H(ID_{HA}||IP_{HA})$. 
\begin{figure}[H]
	\centering
	\includegraphics[width=3.2in]{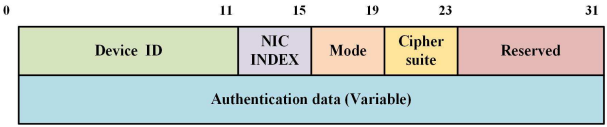}
	\caption{Packet header}
	\label{header}
\end{figure}
We use the pseudo-random function (PRF) to generate the shared key, $k^{IF_{num}} = PRF (K^{IF_{num}}, seed_{initial})$, where $k$ represents the key material and $seed$ represents the number of IP changes of the interface network card. When MR communicates with HA for the first time, the seed counter of each network card is set to 1 (the security can be improved by setting the initial value), and the counter step size increases with each change of the network adapter. Both communication parties shall maintain the same seed counter. To ensure the consistency of shared keys.

For the multihoming scenario, we take the network card identification number of the local device as one of the notification messages, which is helpful for the communication parties to manage the multihoming address. At the same time, the local device ID should be notified to the opposite end to identify the required security domain parameters. The data packet header is shown in Fig \ref{header}. The value of the authentication field is obtained by performing the HMAC operation on the immutable part of the IP packet. The HMAC key is the negotiated shared key to ensure data integrity and source authentication.
\begin{figure}[!t]
	\centering
	\includegraphics[width=3.2in]{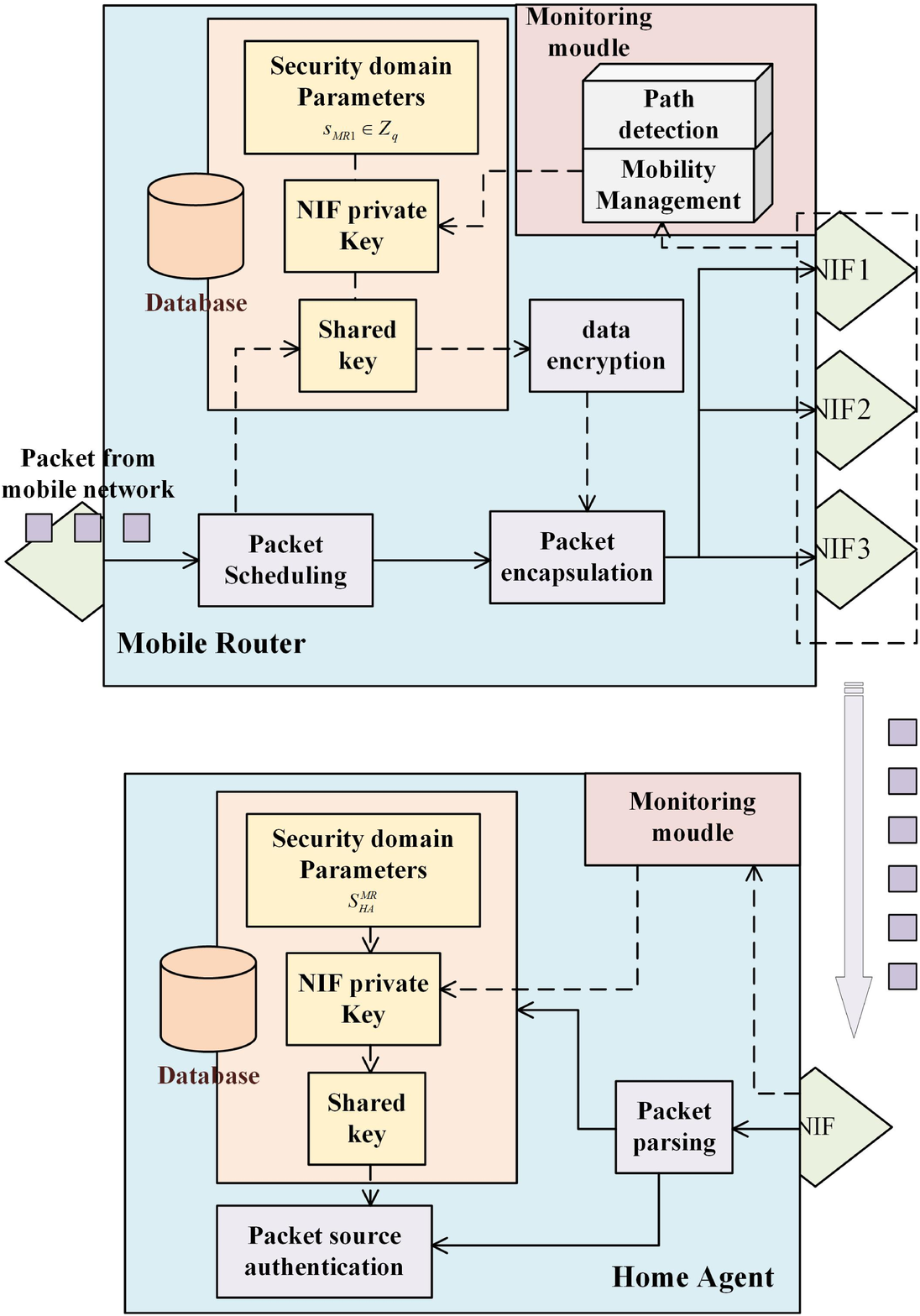}
	\caption{Communication flow}
	\label{fig3.3}
\end{figure}

\subsubsection{HA receive}

When the packet arrives at HA, the packet is parsed first. HA obtains the peer device number, and checks whether there is a corresponding registration item in the security domain database. If so, it indicates that the peer device is legal. It checks the corresponding shared private key according to the peer network card adapter number. If the network card adapter arrives for the first time, it initializes the private key. The shared key material is calculated as $K^{IF_{num}}_{MR}=e(H(ID_{MR}||IP_{MR}^{IF_{num}}||IF_{num}),S_{HA}^{MR})$. Then, the shared private key is extracted by PRF for HMAC calculation and comparison of the corresponding fields of the data packet, so as to realize the integrity and source authentication of the received packet. The complete communication process is shown in Fig \ref{fig3.3}.

It is worth noting that the above authentication information only needs to add some fields in the data packet header. When there is a user data transmission demand in the mobile subnet, the authentication process can be triggered through Mr forwarding data without additional information exchange. It perfectly follows the design principle of the IP layer, that is, only forwarding is realized, and the reliable data transmission is guaranteed by the upper layer.


\subsection{Mobile support}

When the packet arrives at HA, the packet is parsed first. HA obtains the peer device number and checks whether there is a corresponding registration item in the security domain database. If so, it indicates that the peer device is legal. According to the peer network card adapter number, it checks the corresponding shared private key. If the network card adapter arrives first, it initializes the private key. The shared key material is calculated as $K^{IF_{num}}_{MR}=e(H(ID_{MR}||IP_{MR}^{IF_{num}}||IF_{num}),S_{HA}^{MR})$. Then, the shared private key is extracted by PRF for HMAC calculation and comparison of the corresponding fields of the data packet to realize the integrity and source authentication of the received packet. The complete communication process is shown in Fig \ref{fig3.3}.

It is worth noting that the above authentication information only needs to add some fields in the data packet header. When there is a user data transmission demand in the mobile subnet, the authentication process can be triggered through Mr forwarding data without additional information exchange. It perfectly follows the design principle of the IP layer, that is, only forwarding is realized, and the upper layer guarantees reliable data transmission.


According to the communication status between MR and HA, the switching mode of the proposed strategy is divided into two cases. The first case is that the switching occurs during the data communication between both parties. The IP switching information is not sent separately but still waiting for the user data packet to be updated and triggered. We call it the transmission mode. The second situation is that there is no data interaction between the two sides at switching. At this time, MR will actively send address update notification, similar to binding update (BU) information in NEMO, to prevent HA from initiating communication first. We call it notification mode. Even if it is notification mode, only one interaction is required to complete all switching processes, including key renegotiation. For the security problem of mobile multihoming devices (multihoming can be regarded as a particular form of mobile), the traditional security policies of IPSec renegotiate the SA in the form of IP notification to realize state switching, which will lead to the terminal of data transmission in the negotiation process, so it has poor support for mobile scenarios.


\subsection{Key revocation}
When the MR device is no longer trusted (when the device is lost or hijacked), ha deletes the corresponding parameters of the corresponding MR security domain, and MR becomes a registered illegal device.

\section{The Analysis and principle of DDGS scheme}

In this section, to prove the performance of the proposed scheme in the mobile multihoming scenario, we compare it with IKEv2 based on the following performance metrics.

\begin{itemize}
	\item Authentication Latency: The time required for MR to initiate authentication until the authentication is successful and negotiate the shared key.
	\item Handover Latency: The time required for The mobile router updates the binding with the communication peer and negotiates the shared key after the mobile router reacquires the address.
	\item Data latency: Processing delay overhead caused by security mechanism.
	\item Multi-path parallel throughput: The impact of security mechanism on multi-path parallel transmission performance is expressed by aggregating throughput
\end{itemize}

\subsection{Network Topology for Simulation}
\begin{figure}[H]
	\centering
	\includegraphics[width=3.5in]{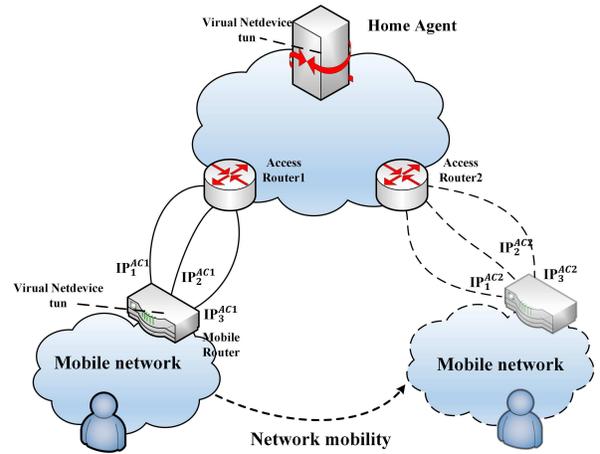}
	\caption{Simulation topology}
	\label{simulation}
\end{figure}

To check the performance difference between the proposed strategy and IKEv2 in different network states under heterogeneous network scenarios, we designed the simulation topology as shown in Fig. \ref{simulation}. We define a mobile network with multiple network interfaces and define two wireless access points. Through network segment planning, the mobile router can switch between access points. We establish a tunnel between the mobile router and HA as a virtual network device and establish a multi-link aggregation scheduling mechanism in the tunnel. The scheduling strategy is EDPF, and the user communication data flow will be aggregated and transmitted in the tunnel. The code construction of our security mechanism is based on the Pairing-Based Cryptography (PBC) library. The PBC library is a free C library GMP library that performs the mathematical operations underlying pairing-based cryptosystems\cite{TP-toolbox-web}. The pseudo-code of the C++ program we built is as above.

\begin{breakablealgorithm}
	\caption{MR Sender}
	\begin{algorithmic}[1]
		\Function {loadconfig ()}{}  
		\State System\_Setup ($pairing$, $s_{MR}$) 
		\State Global\_param.CalculPk\_HA($IP_{HA}$, $ID_{HA}$)
		\EndFunction  
		\Function {Netmonitor ()}{} 
		\While {$true$}
		\If {adapterstatus==adapterinit}
		\State Upadapter.CalculSk\_MR($this->index$, $IP_{MR}$, $ID_{MR}$)
		\State Upadapter.Calculsharekey($this->Sk\_MR$, $Global\_param.Pk\_HA$)
		\EndIf
		\If {adapterstatus==adapterreload}
		\State Upadapter.UpdateSk\_MR($newIP_{MR}$, $ID_{MR}$, $this->index$)
		\State Upadapter.Updatesharekey($this->Sk\_MR$, $Global\_param.Pk\_HA$)
		\EndIf
		\EndWhile
		\EndFunction
		\Function {Datatrans ()}{} 
		\While {recvfrom(tunadapter)}
		\State Schedule()
		\State Upadapter.packetAUTH()
		\State Upadapter.send()
		\EndWhile
		\EndFunction
		\label{code:recentEnd}
	\end{algorithmic}
\end{breakablealgorithm}

\begin{breakablealgorithm}
	\caption{HA Receiver}
	\begin{algorithmic}[1]
		\Function {loadconfig ()}{}  
		\State System\_Setup ($pairing$, $s_{MR}$) 
		\EndFunction  
		\Function {Datatrans ()}{} 
		\While {recvfrom(Upadapter)}
		\State Parsingpacket()
		\If {packet.MRID==REGISTERED}
		\If {packet.MRID==LOADED}
		\If {packet.adapterIndex==LOADED}
		\If {packet.ip==OLD\_IP}
		\If {Upadapter.verify()==ture}
		\State packet.recv()
		\Else 
		\State packet.drop()
		\EndIf
		\Else	
		\State Upadapter.Updatesharekey($new$ $IP_{MR}$)
		\EndIf
		\Else	
		\State Upadapter.loadadapterIndex($IP_{MR}$, $index$)
		\EndIf
		\Else	
		\State Upadapter.loadMR($IP_{MR}$, $ID_{MR}$, $index$)
		\EndIf
		\Else
		\State packet.drop()
		\EndIf
		\EndWhile
		\EndFunction
		\label{code:recentEnd}
	\end{algorithmic}
\end{breakablealgorithm}
\begin{figure*}[tp]
	\centering
	
	\subfigtopskip=2pt 
	\subfigbottomskip=2pt 
	\subfigcapskip=-5pt 
	
	\subfigure[0\% packet loss rate]{
		\includegraphics[scale=0.30]{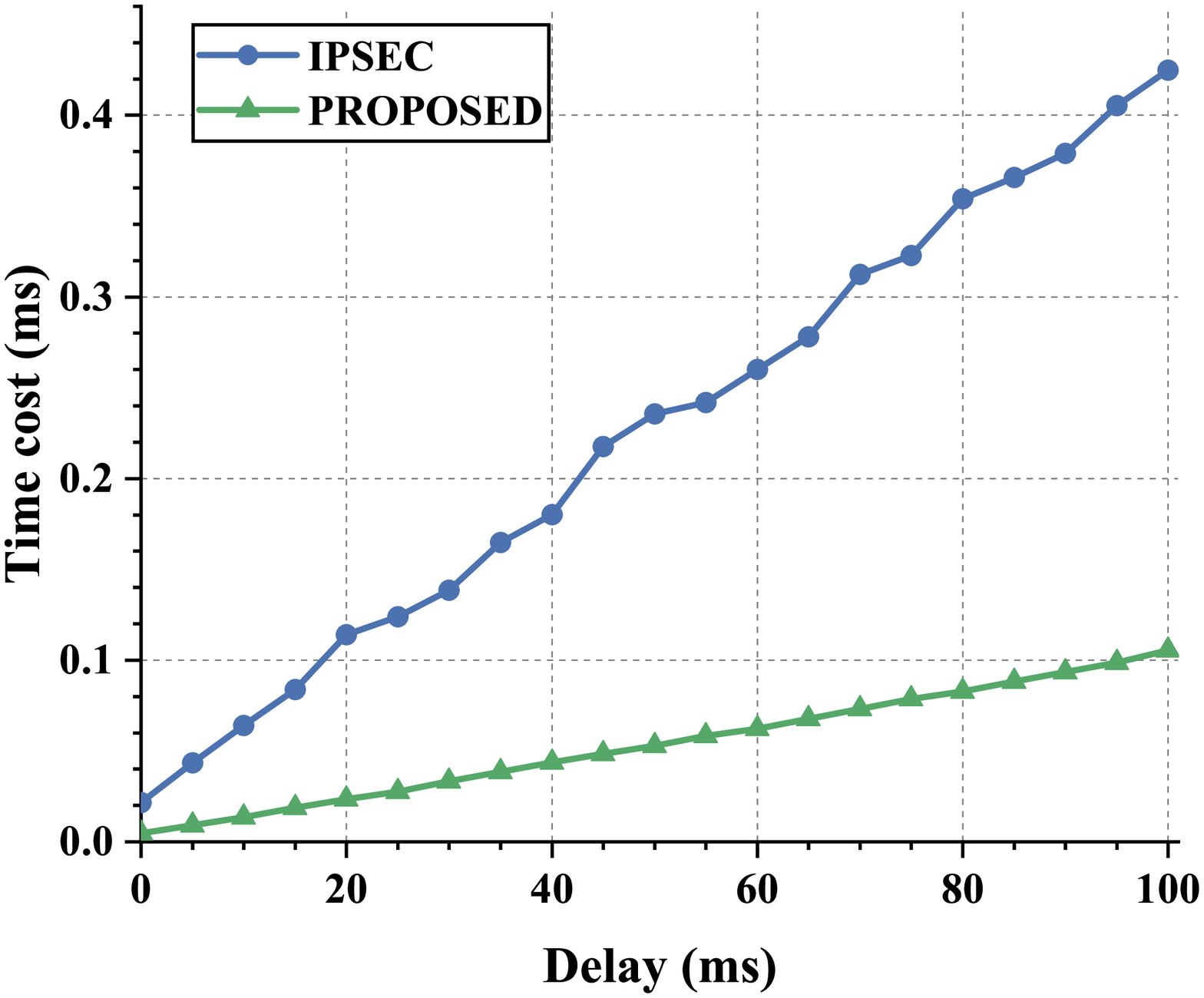}}
	\hspace{0.3in} 
	\subfigure[5\% packet loss rates]{
		\includegraphics[scale=0.30]{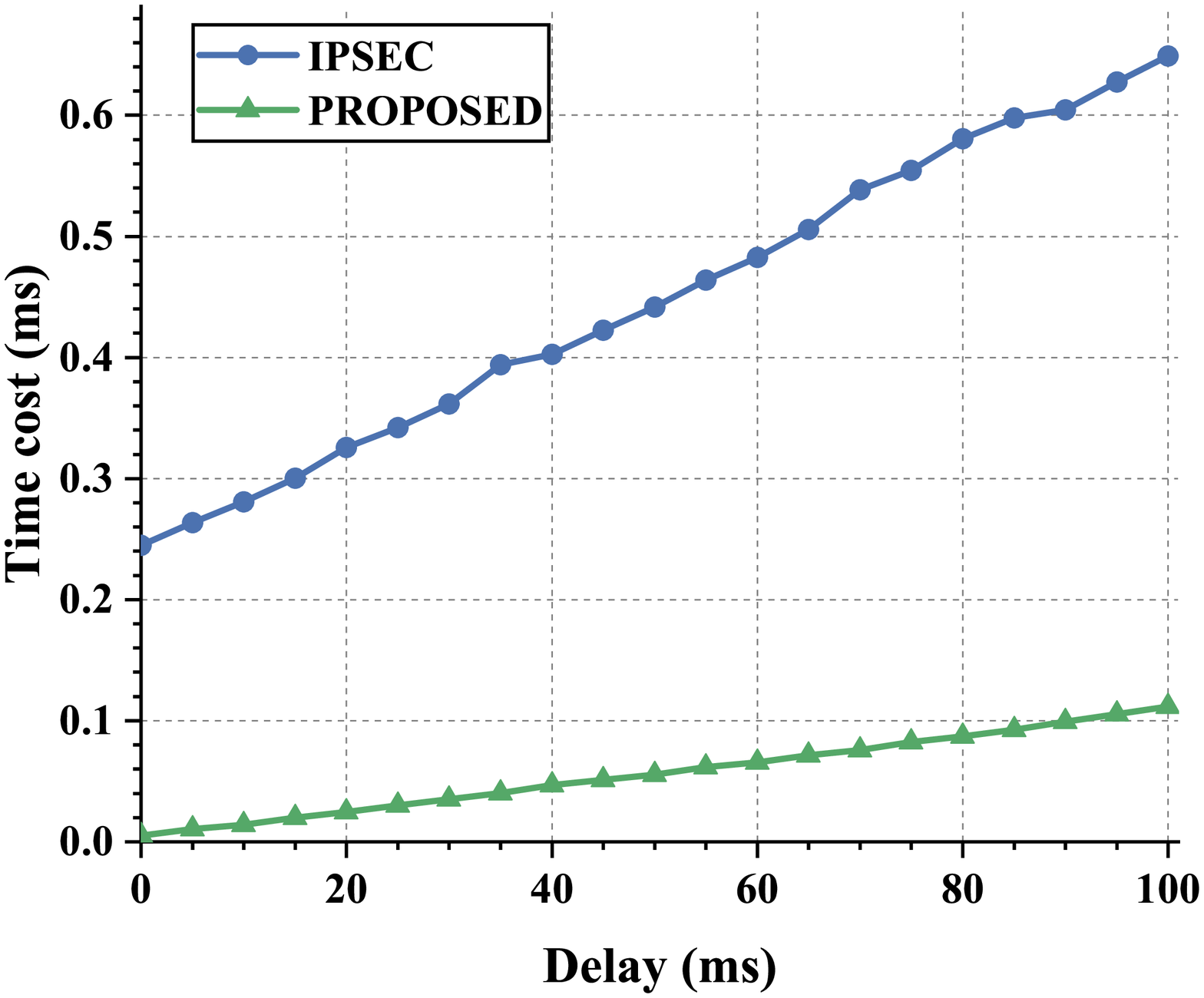}}

	\subfigure[10\% packet loss rate]{
		\includegraphics[scale=0.30]{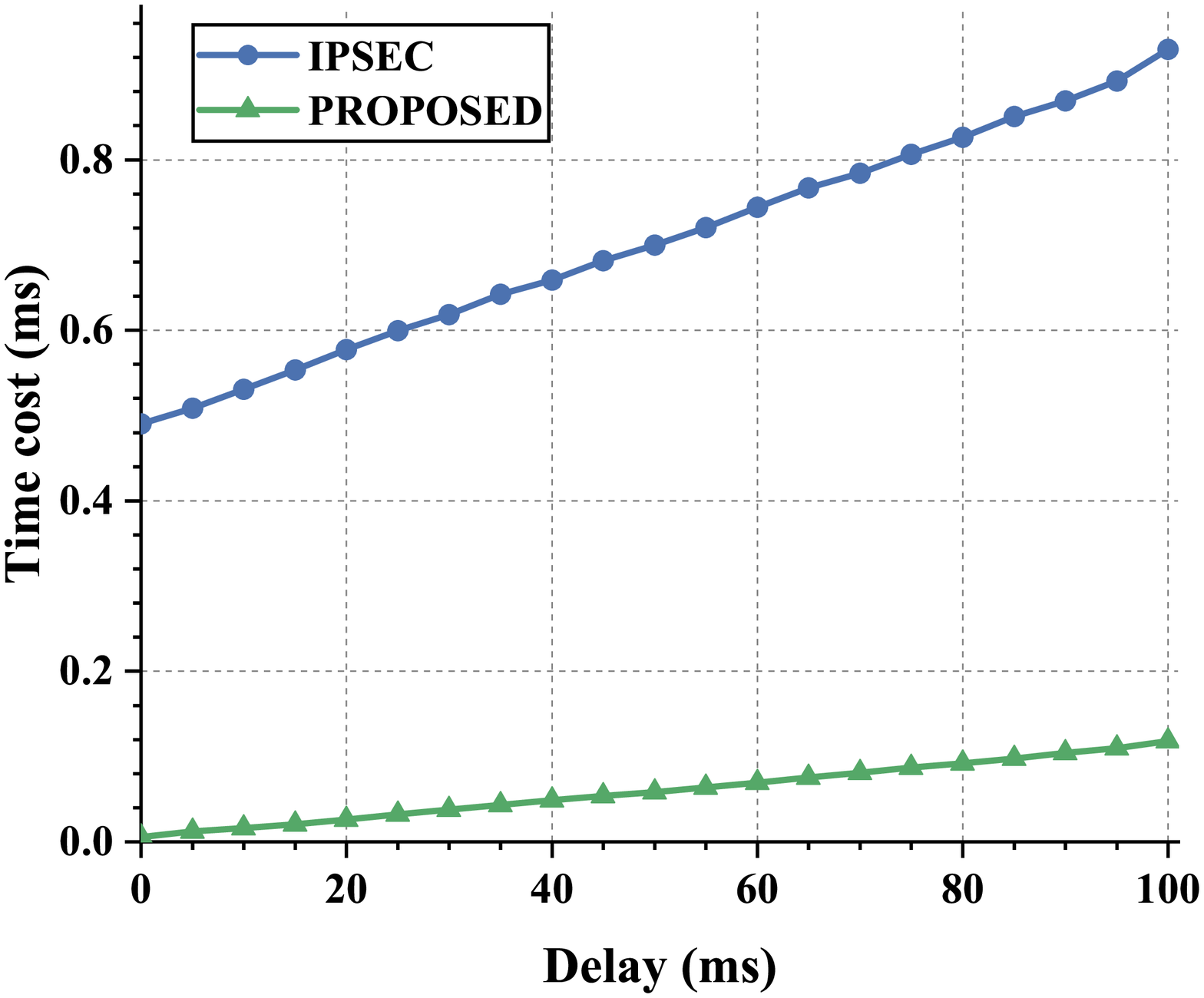}}
	\hspace{0.3in} 
	\subfigure[15\% packet loss rate]{
		\includegraphics[scale=0.30]{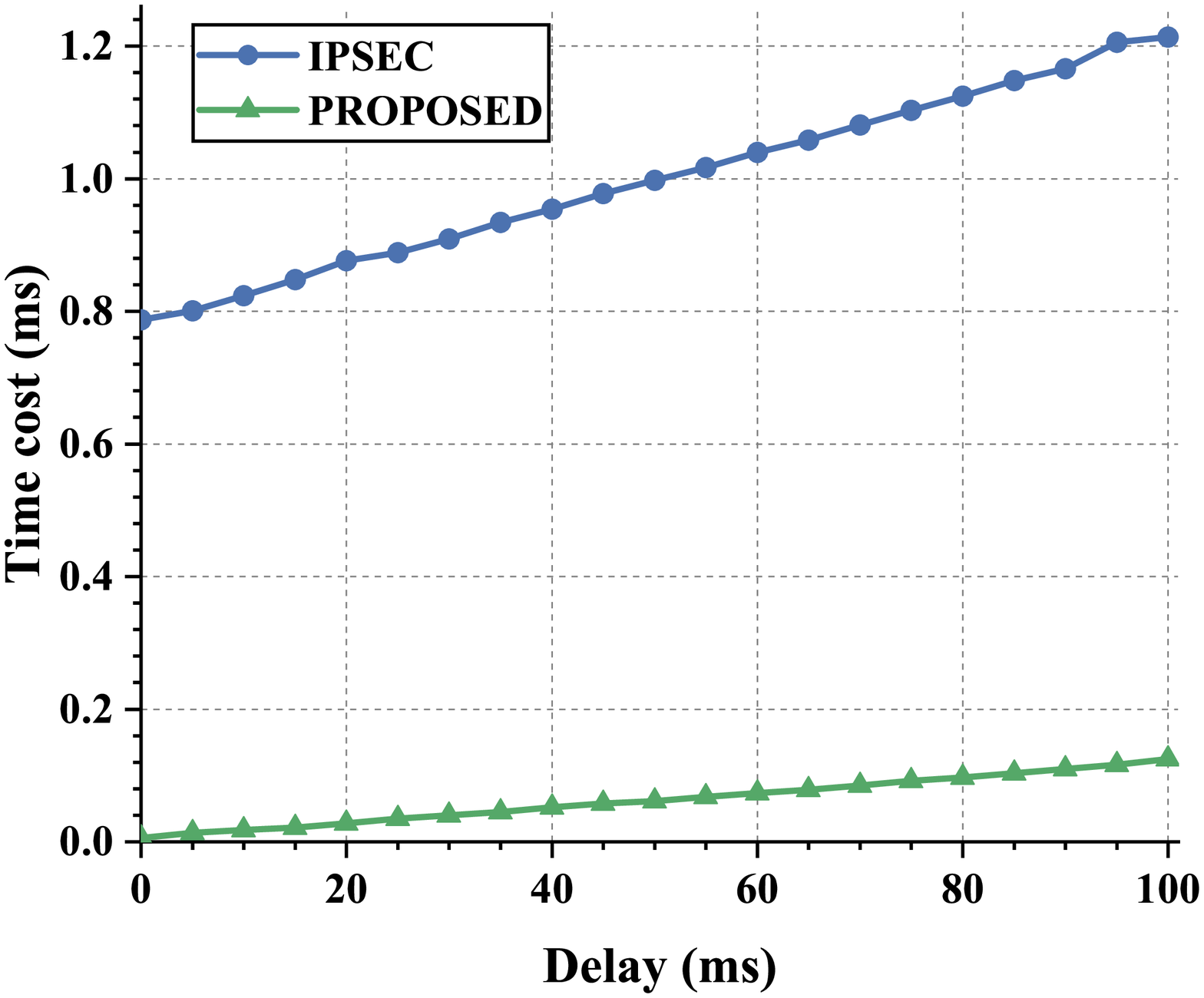}}
	\caption{The authentication latency under different packet loss rate scenarios}
	\label{fig5}
	
\end{figure*}

\begin{figure*}[tp]
	
	\centering
	
	\subfigtopskip=2pt 
	\subfigbottomskip=2pt 
	\subfigcapskip=-5pt 

	\subfigure[0\% packet loss rate]{
		
		\includegraphics[scale=0.30]{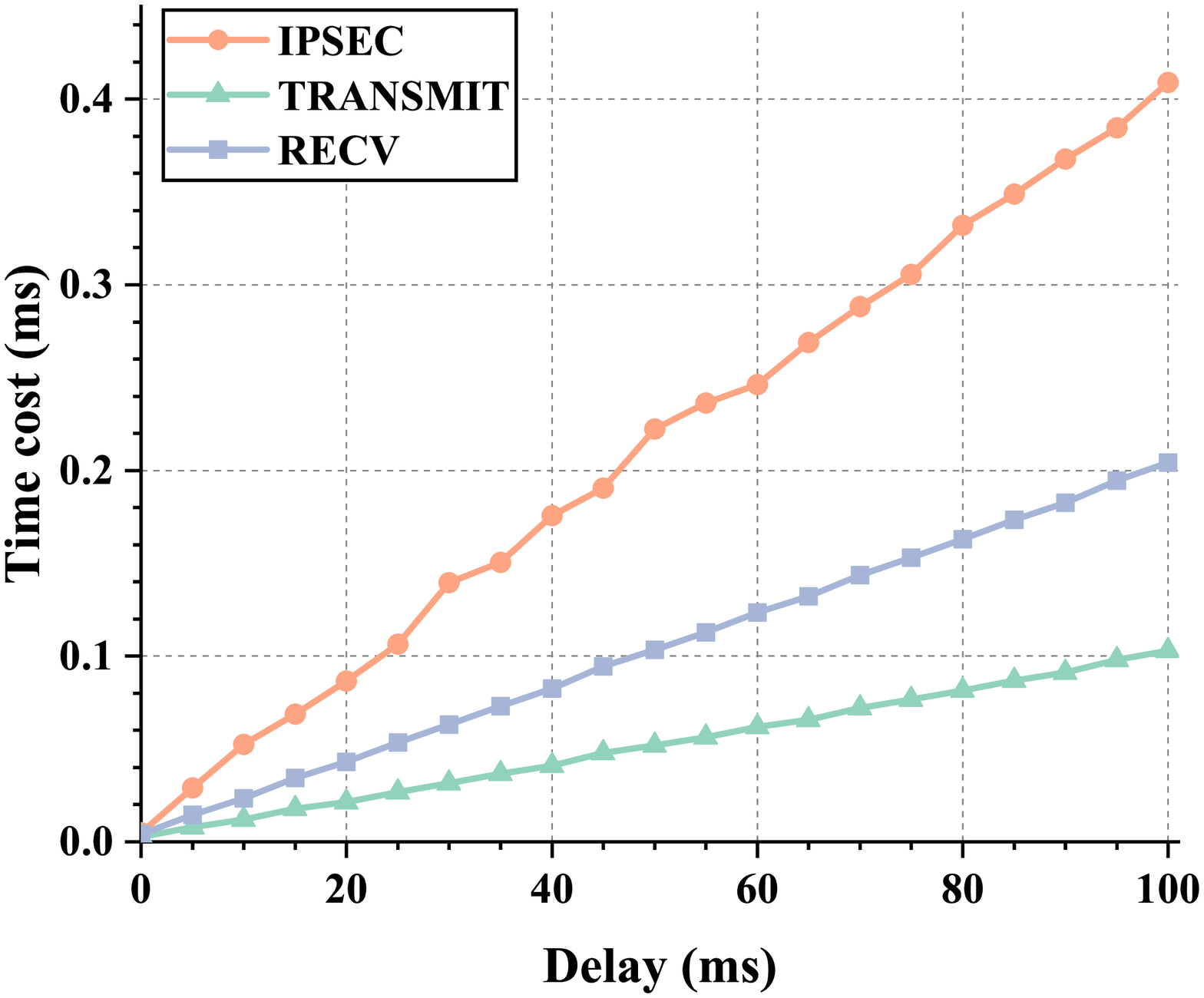}}
	\hspace{0.3in} 
	\subfigure[5\% packet loss rate]{
		
		\includegraphics[scale=0.30]{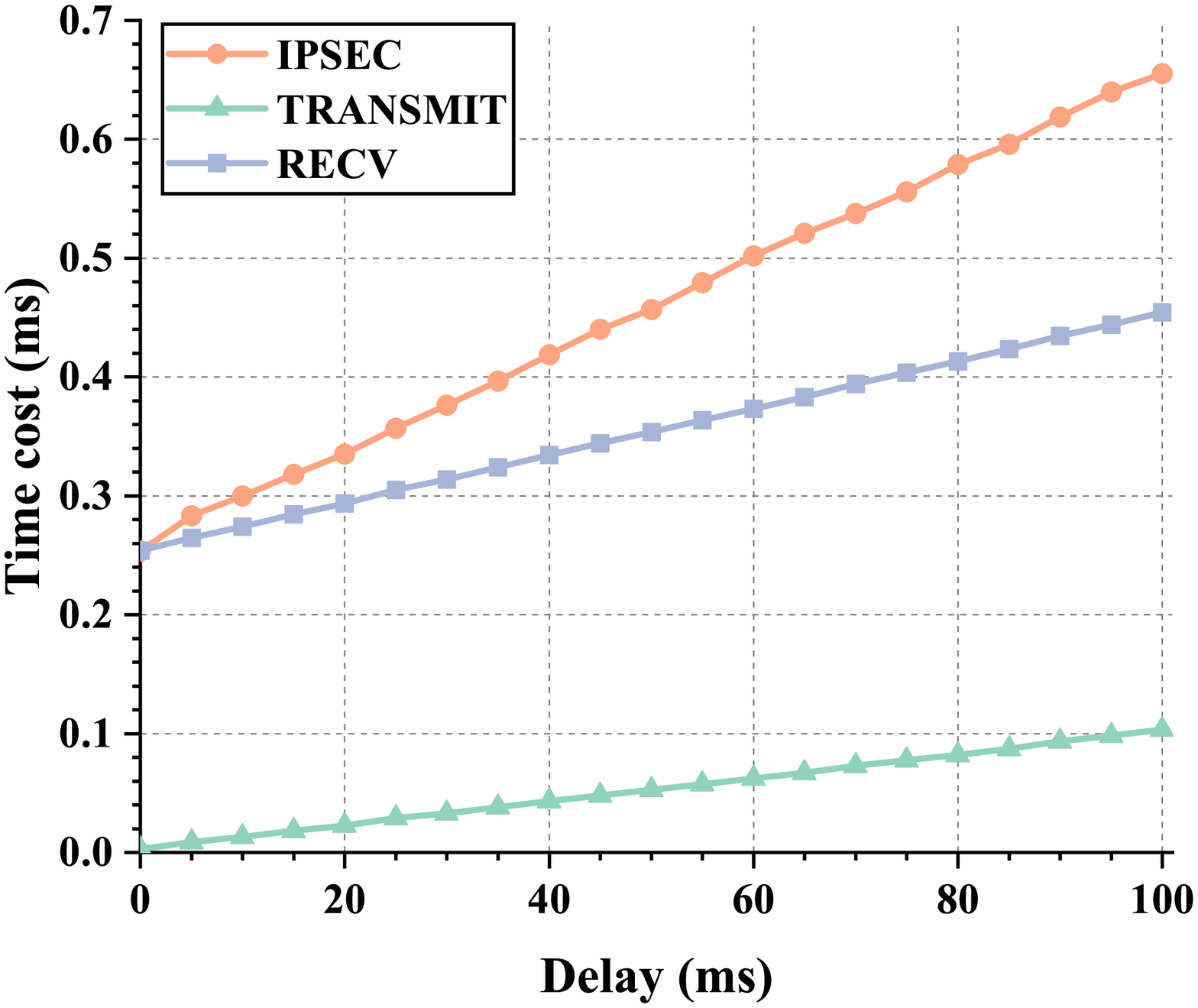}}

	\subfigure[10\% packet loss rate]{
		
		\includegraphics[scale=0.30]{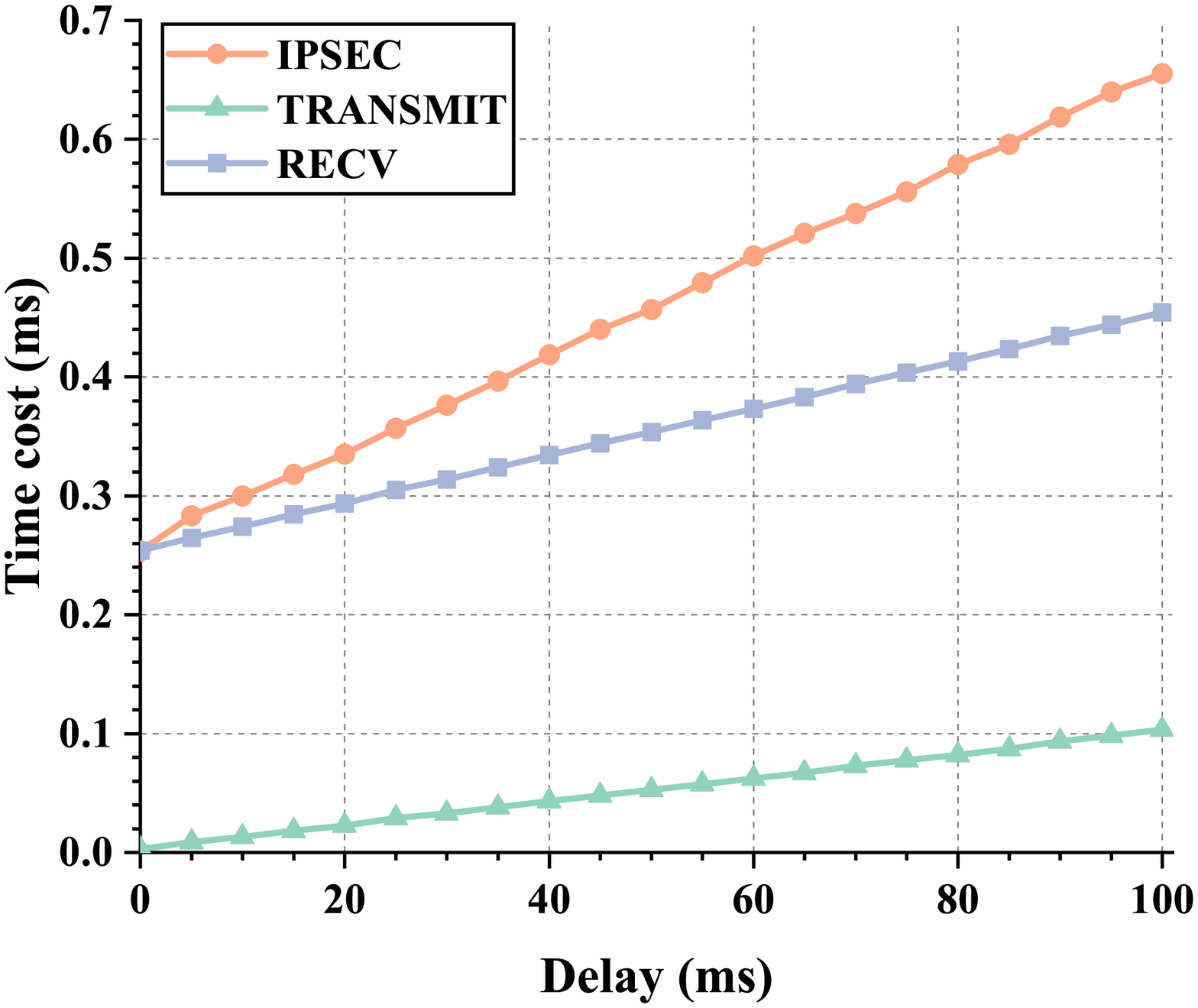}}
	\hspace{0.3in} 
	\subfigure[15\% packet loss rate]{
		
		\includegraphics[scale=0.30]{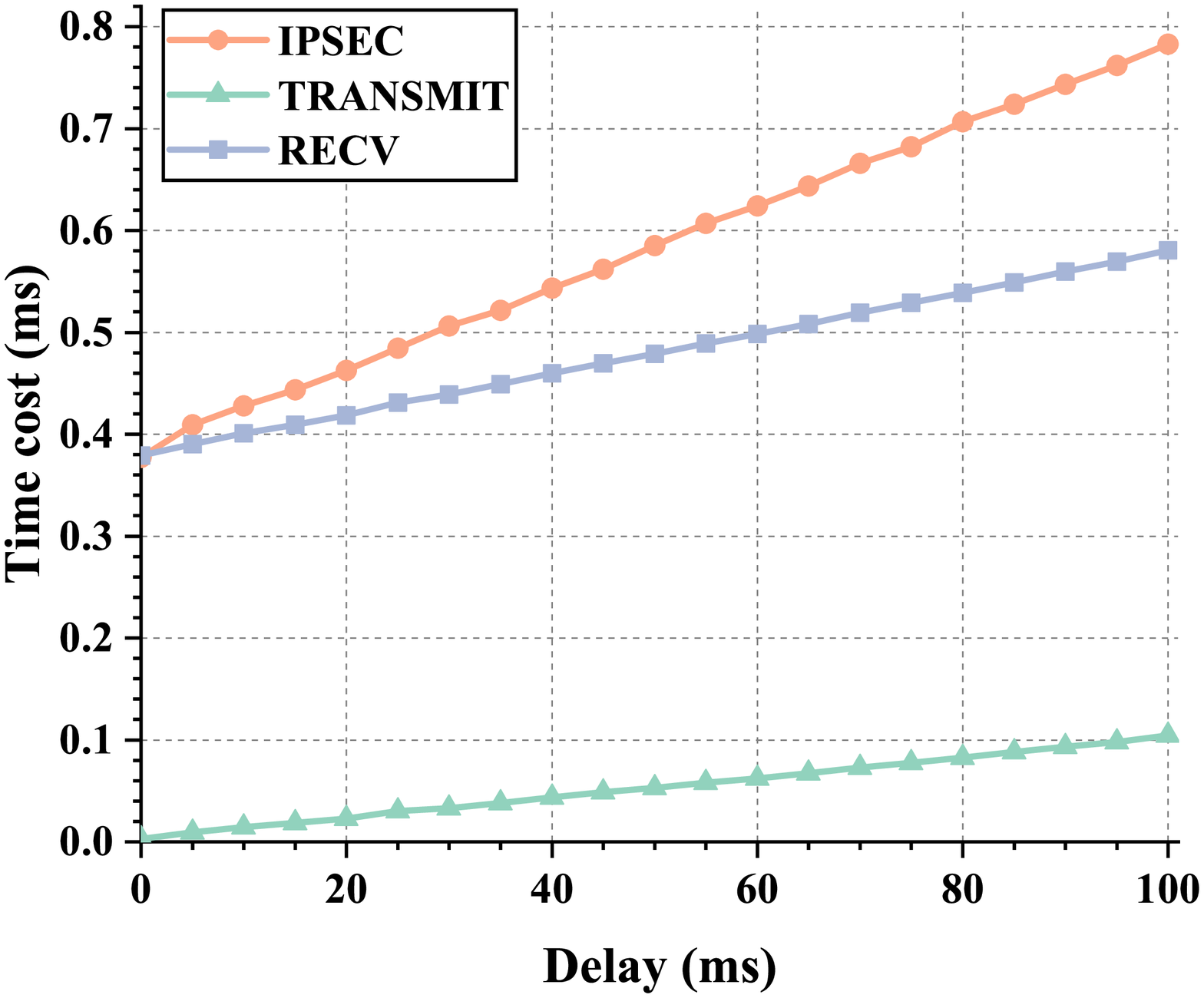}}
	\caption{The handover latency under different packet loss rate scenarios}
	\label{fig5.3}
	
\end{figure*}

\subsection{Authentication Latency}


The authentication delay is designed as the time required for MR to receive subnet data, with data communication requirements to trigger the authentication process until both parties can normally carry out data communication. Because the working scenario of the proposed security mechanism is a heterogeneous wireless mobile network, it has the characteristics of a high packet loss rate and large delay fluctuation. We tested the impact of link delay on authentication delay under different link packet loss rates. We tested the authentication processes of different strategies many times and averaged them. The results shown in Figure 5 show that the proposed mechanism still has a short authentication delay under the scenario of a high packet loss rate and high delay. The delay reduction brought by the proposed strategy comes from two aspects. 1. The proposed strategy is based on the data interaction establishment of the network layer, and there is no need to establish a stateful security alliance similar to IPSec, which can avoid the packet loss caused by the state establishment signaling in the scenario of high packet loss rate. 2. the non-interactive key agreement only depends on the necessary relevant information in data transmission, such as an address, which reduces the transmission delay caused by interactive negotiation, especially the overhead in high delay scenarios.

\subsection{Handover Latency}

This paper only aims at the IP address switching caused by the movement of the mobile router. Because the IP address is the state parameter of SA, the purpose of the IKEv2 mobile multihoming extension mobike is to expand the IP support of the stateful security alliance. To realize the state update, it is necessary to update the notification and renegotiate the key for the new state. This paper compares it with the two switching modes of the proposed strategy. It can be seen from Figure 6 that in the data transmission mode, there is no need to confirm the return state, which retains the stateless transmission characteristics of the IP layer. The reliable transmission is guaranteed through the upper layer protocol, which improves the switching efficiency. Due to the non-interactive key agreement, the no-load mode still has a smaller switching delay compared with mobike.

\subsection{Data lantency}


This part discusses the delay impact of the designed security mechanism in the multi-path transmission scenario. In the simulation, we negotiate the IPSec ah protocol as sha256 and use the same data authentication algorithm in the proposed strategy to compare the performance. 

The multi-link state parameters are shown in Figure XX to simulate the heterogeneous scenario of multi-link. The aggregation scheduling algorithm can realize the optimal scheduling of data delay.

\begin{figure}[H]
	\centering
	\includegraphics[width=3.8in]{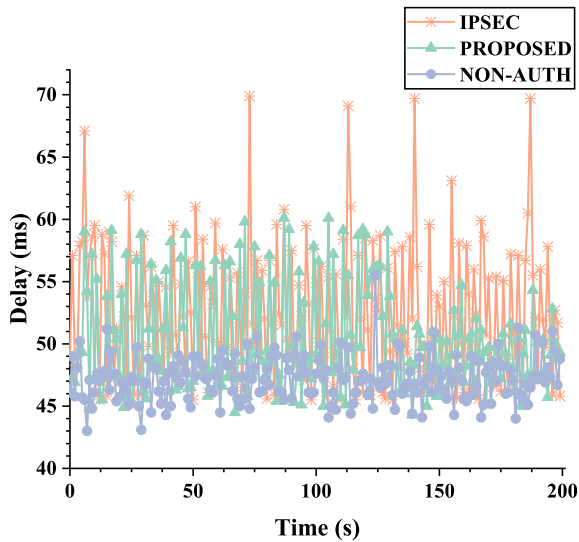}
	\caption{Data transmission lantency} 
	\label{fig5.5}
\end{figure}

\begin{table}[htbp] 
	\centering
	\caption{Links Configurations}
	\label{table5.2}
	
	\setlength{\tabcolsep}{2mm}{
		\begin{tabular}{cccc} 
			\toprule 
			parameters & Link A & Link B & Link C \\ 
			\midrule 
			loss rate $\left( \% \right)$ & 1.5\textasciitilde2 & 2\textasciitilde2.5 & 1\textasciitilde1.5 \\ 
			bandwidth & 4 Mbps & 4 Mbps & 4 Mbps \\ 
			delay & 40\textasciitilde50 ms & 60\textasciitilde70 ms & 50\textasciitilde60 ms \\ 
			\bottomrule

	\end{tabular} }
	
\end{table}

We send ICMP detection packets to the opposite virtual network port address through the tunnel virtual network port address at the MR end at an interval of seconds and record the proposed strategy. The multihoming network port establishes Ike SA with the opposite end, respectively, and the delay information under no load. As shown in the figure, compared with no load, the proposed scheme and IPSec bring unavoidable additional delay overhead caused by data processing and verification between devices. Because the multipath environment amplifies the impact of time delay on transmission, it can be concluded that the overall delay brought by the proposed strategy is smaller than that brought by IPSec, And the fluctuation is low.

\subsection{Throughput}

\begin{figure}[H]
	\centering
	\includegraphics[width=3.8in]{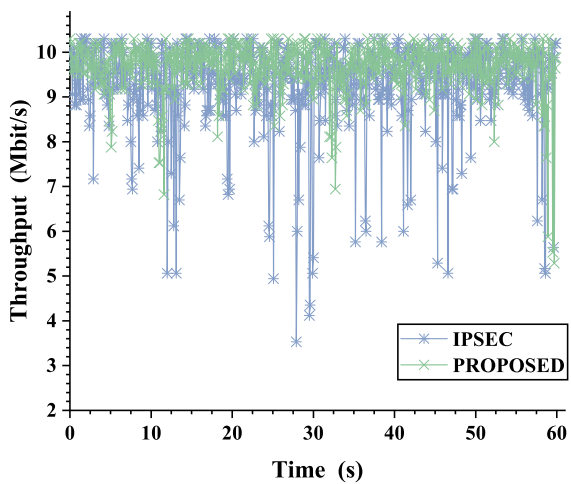}
	\caption{Bandwidth aggregation efficiency} 
	\label{fig5.5}
\end{figure}

Finally, we evaluate the throughput of the proposed security policy in the multi-link scenario, which is also the most key parameter of the multi-link parallel transmission system. The link settings are shown in Table XX. To evaluate the maximum load of the system, UDP tests are conducted on both sides of the communication through iperf to evaluate the load performance of the system. Our results show that the maximum load of the system is 10m, which is tested as data traffic. The results are shown in Figure XX. It can be seen that the proposed strategy has better multi-link aggregation performance than IPSec. Because the additional delay overhead brought by security measures will affect the transmission of packets and reduce the aggregation efficiency, the proposed strategy has less bandwidth aggregation efficiency degradation caused by packet disorder.

\section{Conclusion}
This paper aims at designing a lightweight security authentication scheme for mobile and multi-homing devices working in wireless heterogeneous network scenarios. We proposed the NIMSA scheme. Due to identity-based non-interactive key exchange characteristics, our proposed scheme can significantly reduce the delay and bandwidth overhead. Through experimental comparison with IKEv2 and its mobile extension, we fully prove the advantages of the proposed strategy compared with the traditional scheme.

%



\section*{Acknowledgment}

This work was supported in part by the National Natural Science Foundation of China (NSFC) under Grant 61872029, and in part by the Beijing Municipal Natural Science Foundation under Grant 4182048.

\ifCLASSOPTIONcaptionsoff
  \newpage
\fi



%

\bibliographystyle{./Transactions-Bibliography/IEEEtran}
\bibliography{./Transactions-Bibliography/IEEEabrv,./bib/papernote}
%




\end{document}